\documentclass[showkeys,reprint,amsmath,amssymb,aps,superscriptaddress]{revtex4-2}
\usepackage[T1]{fontenc}
\usepackage[utf8]{inputenc}
\usepackage{graphicx}
\usepackage[]{subfig}

\begin{document}
\title{A threshold model of plastic waste fragmentation: New insights
  into the distribution of microplastics in the ocean and its
  evolution over time} \author{Matthieu \surname{George}}
\affiliation{Laboratoire Charles-Coulomb, UMR 5221 CNRS -- université
  de Montpellier, Campus Triolet, Place Eugène-Bataillon -- CC069,
  F-34095 Montpellier Cedex 5 -- FRANCE} \author{Frédéric
  \surname{Nallet}} \affiliation{Centre de recherche Paul-Pascal, UMR
  5031 CNRS -- université de Bordeaux, 115 avenue du
  Docteur-Schweitzer, F-33600 Pessac -- FRANCE} \author{Pascale
  \surname{Fabre}} \email[Email for correspondence:
]{pascale.fabre@umontpellier.fr} \affiliation{Laboratoire
  Charles-Coulomb, UMR 5221 CNRS -- université de Montpellier, Campus
  Triolet, Place Eugène-Bataillon -- CC069, F-34095 Montpellier Cedex
  5 -- FRANCE}
\date{\today}
\begin{abstract}
Plastic pollution in the aquatic environment has been assessed for
many years by ocean waste collection expeditions around the globe or
by river sampling. While the total amount of plastic produced
worldwide is well documented, the amount of plastic found in the
ocean, the distribution of particles on its surface and its evolution
over time are still the subject of much debate. In this article, we
propose a general fragmentation model, postulating the existence of a
critical size below which particle fragmentation becomes extremely
unlikely. In the frame of this model, an abundance peak appears for
sizes around 1mm, in agreement with real environmental data. Using, in
addition, a realistic exponential waste feed to the ocean, we discuss
the relative impact of fragmentation and feed rates, and the temporal
evolution of microplastics (MP) distribution. New conclusions on the
temporal trend of MP pollution are drawn.
\end{abstract}
\maketitle
\section{Introduction}
\label{intro}
Plastic waste has been dumped into the environment for nearly 70
years, and more and more data are being collected in order to quantify
the extent of this pollution. Under the action of degradation agents
(UV, water, stress), plastic breaks down into smaller pieces that
gradually invade all marine compartments. If the plastic pollution
awareness initially stemmed from the ubiquitous presence of
macro-waste, it has now become clear that the most problematic
pollution is ``invisible'' \emph{i.e.} due to smaller size debris, and
the literature exploring microplastics (MP, size between 1~$\mu$m and
5~mm) and nanoplastics (NP, size below 1~$\mu$m) quantities and
effects is rapidly increasing. The toxicity of plastic particles being
dependent on their sizes and concentrations, it is crucial to know
these two parameters in the natural environment to better predict
their impacts.  While the total amount of plastic produced worldwide
is well-documented~\cite{ourWorldInData}, the total amount of plastic
found in the ocean and its time evolution are still under debate:
while many repeated surveys and monitoring efforts have failed to
demonstrate any convincing temporal trend~\cite{galgani2021},
increasing amounts of plastic are found in some regions, especially in
remote areas, and a global increase from \emph{ca.} 2005 has been
suggested~\cite{eriksen2023}. In terms of particle sizes, a lot of
efforts have been devoted to collect and accurately identify smaller
and smaller MP, leading to the conclusion that MP smaller than
\emph{ca.} 300~$\mu$m were more numerous than larger
MP~\cite{Poulain2019,Pabortsava2020,Weiss2021}.  \par The purpose and
originality of this paper is to explain in a global manner the
observed size distribution of particles floating in the
oceans. Indeed, this distribution exhibits, in our opinion, some very
specific features that were pointed out in a previous
paper~\cite{george2021}. It is important to note that these features
are common to all papers reporting particle size
data~\cite{Poulain2019,cozar2014plastic,isobe2014,Eriksen2014,ter2016understanding,Lebreton2018,Lebreton2019,Isobe2019,Kataoka2019,Lindeque2020,Pabortsava2020,Tokai2021,Weiss2021}:
when browsing the sizes from the largest to the smallest the particle
abundance increases, first following closely a power-law, but the
increase in the number of particles then slows down and an abundance
(broad) peak is observed around
1~mm~\cite{cozar2014plastic,ter2016understanding,Lebreton2018}. Between
1~mm and approximately 150~$\mu$m, the number of particles is reported
to \emph{decrease}. The abundance increases again from approximately
150~$\mu$m down to 10~$\mu$m, with an amount of small MP which is
about two orders of magnitude larger than what is found around
1~mm~\cite{Poulain2019,Pabortsava2020}. While such a description of
the available data may appear as qualitative only, owing to the
difficulty of comparing data for large and small MP collected over the
years in different places and with different methods, we give in
Section~\ref{ResultsDiscussion} below arguments supporting our
analysis in quantitative terms.  \par To the best of our knowledge,
the physical reason~\cite{andrady2017,george2021} for the existence of
two rather different size distributions for microplastics (small MP
$<150$~$\mu$m, large MP between 1 and 5~mm) is that there are two
rather different fragmentation pathways: \emph{i)} bulk fragmentation
with iterative splitting of one piece into two daughters for large MP,
and \emph{ii)} delamination and disintegration of a thin surface layer
(around 100~$\mu$m depth) into many smaller particles for small
MP. This description does however not explain the deficit of
microplastics in the size range extending from $\approx150$~$\mu$m to
1~mm.  Early authors attempted to describe the large MP distribution
by invoking a simple iterative fragmentation of plastic pieces into
smaller objects, conserving the total plastic
mass~\cite{cozar2014plastic,Eriksen2014,Cozar2015,enders2015abundance,van2015global,ter2016understanding,Lebreton2019,Tokai2021},
in accordance to pathway \emph{i)}. These models lead to a
time-invariant power-law dependence of the MP abundance with size
(refer to Supplementary Information~\ref{standardModel} for an
elementary version of such models), which is in fair agreement with
experimental observations for large MP. However, they fail to describe
the occurrence of an abundance peak and the subsequent decrease of the
number of MP when going to smaller sizes. Various mechanisms such as
sinking to the ocean floor, ingestion by marine organisms, mesh size
of the collecting devices, etc. have been invoked to qualitatively
explain the absence of particles smaller than 1~mm. Very recently, two
papers have addressed this issue using arguments related to the
fragmentation process itself. Considering the mechanical properties of
a one-dimensional material (flexible and brittle fibers) submitted to
controlled stresses in laboratory mimicking ocean turbulent flow,
Brouzet \emph{et al}~\cite{Brouzet2021} have shown both theoretically
and experimentally in the one-dimensional case that smaller pieces are
less likely to break. Aoki and Furue~\cite{Aoki2021} reached
theoretically the same conclusion in a two-dimensional case using a
statistical mechanics model. Note that both approaches are based on
the classical theory of rupture, insofar as plastics fragmenting at
sea have generally been made brittle by a long exposure to UVs.  \par
In this paper, we only explore the fragmentation pathway \emph{i)},
and ignore the delamination pathway \emph{ii)}, since the delamination
process produces directly very small plastic pieces (below typically
$150$~$\mu$m ~\cite{andrady2017,george2021}). Regardless of the
fracture mechanics details \emph{i.e.} the specific characteristics of
the plastic waste (shape, elastic moduli, aging behavior) and the
exerted stresses, we postulate the existence of a critical size below
which bulk fragmentation becomes extremely unlikely. Since many of the
microplastics recovered from the surface of the ocean are film-like
objects (two dimensions exceeding by a large margin the third one)
like those coming from packaging, we construct the particle size
distribution over time based on the very idea of a universal failure
threshold for breaking two-dimensional objects. A very simple
hand-waving argument from every day's life that illustrates this
breaking threshold, is that the smaller a parallelepipedic piece of
sugar is, the harder it is to break it, hence the nickname \emph{sugar
lump} model used in this paper.  Unlike many previous models, which
made the implicit assumption of a \emph{stationary} distribution, we
explicitly described the temporal evolution of the large MP quantity
(see Sections~\ref{fragModelThreshold} and \ref{suppInfo}). Moreover,
injecting a realistic waste feed into the model, we discussed the
synergistic effect of feeding and fragmentation rates on the large MP
distribution, in particular in terms of evolution with time, and
compared to the field data in Section~\ref{ResultsDiscussion}.

\section{Fragmentation model with threshold}
\label{fragModelThreshold}
The \emph{sugar lump} iterative model implemented the two following
essential features: a size-biased probability of fragmentation on the
one hand, and a controlled waste feed rate on the other
hand. Initially, a constant feeding rate was used in the model. In a
second step, the more realistic assumption of an exponentially growing
feeding rate was introduced and discussed in comparison with field data
(See Section~\ref{ExpGrowth}).
 
At each iteration, we assumed that the ocean was fed with a given amount
of large parallelepipedic fragments of length $L_{\mathrm{init}}$,
width $\ell_{\mathrm{init}}$ and thickness $h$, where $h$ is much
smaller than the other two dimensions and length $L_{\mathrm{init}}$
is, by convention, larger than width $\ell_{\mathrm{init}}$.  At each
time step, every fragment potentially broke into two parallelepipedic
pieces of unchanged thickness $h$. The total volume (or mass) was kept
invariant during the process. In addition, we assumed that, if the
fragment ever broke during a given step, it always broke
perpendicular to its largest dimension $L$: A fragment of dimensions
($L$, $\ell$, $h$) thus produced two fragments of respective
dimensions ($\rho L,\ell,h$) and ($\left[1-\rho\right]L,\ell,h$),
$\rho$ being in our model a random number between 0 and 0.5. Note
that, depending on the initial values of $L,\ell$ and $\rho$, one or
both of the new dimensions $\rho L$ and $\left[1-\rho\right]L$ might
become smaller than the previous intermediate size $\ell$: the
fragmentation of a film-like object, at contrast to the case of a
fiber-like object, is not conservative in terms of its largest
dimension~\cite{ter2016understanding,Brouzet2021}. Furthermore, in
order to ensure that the fragment thickness $h$ remained (nearly)
constant all along the fragmentation process, $\rho$ values leading to
$\rho L$ or $\left(1-\rho\right)L$ significantly less than $h$ were
rejected in the simulation. This obviously introduced a short length
scale cutoff, in the order of $h$, and a limiting, nearly cubic, shape
for the smaller fragments (an ``atomic limit'', according to the
ancient Greek meaning).
\begin{figure}[!ht]
\centering
\includegraphics[width=0.7\columnwidth]{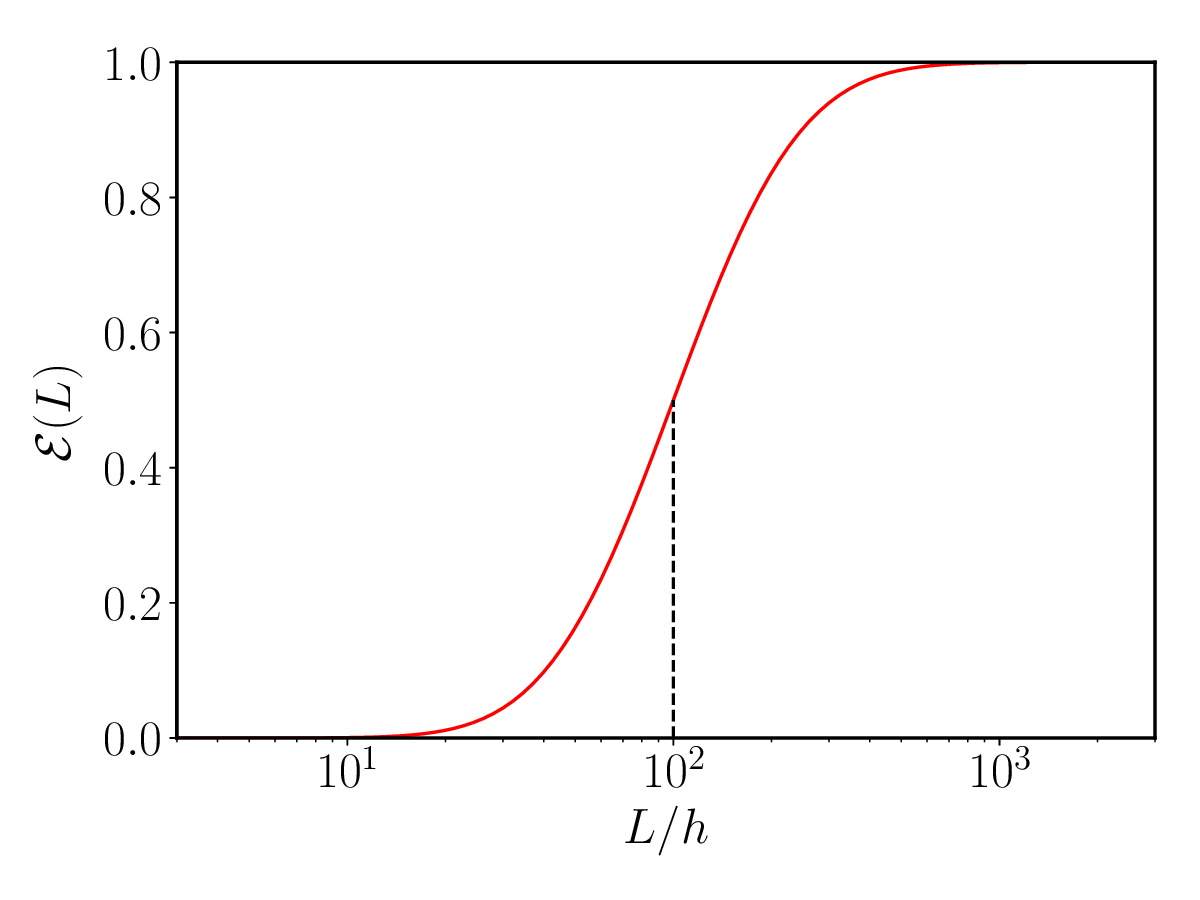}
\caption{Breaking probability of a fragment as a function of the
  largest dimension $L$. The vertical dashed line indicates the
  threshold, here chosen as $L_c=100h$}
\label{fig:efficiency}
\end{figure}

A second length scale, $L_c$, also entered the present model,
originating in the mechanical \emph{sugar lump} approach, described
heuristically by means of a breaking efficiency ${\cal E}(L)$
sigmoidal in $L$. For the sake of convenience, this efficiency was
built here from the classical Gauss error function. It was therefore
close to 1 above a threshold value $L_c$ (chosen large enough compared
to $h$) and close to 0 below $L_c$. A representative example is shown
in Fig.~\ref{fig:efficiency}, with $L_c/h=100$. Note that throughout
this paper, all lengths involved in the numerical model are scaled
by the thickness $h$.
\par Qualitatively speaking, this feature of the model means that when
the larger dimension $L$ is below the threshold value $L_c$, fragments
will ``almost never'' break, even if they have not reached yet the
limiting (approximately) cubic shape of fragments of size $\approx
h$. For the sake of simplicity, the threshold value was assumed
\emph{not} to depend on plastic type or on residence time in the
ocean, considering that weathering occurred from the moment the waste was
thrown in the environment and quickly rendered all common plastics
brittle. A unique $L_c$ was thus used for all fragments.  
\par Technical
details about the model are given in supplementary
information~\ref{suppInfo}.
\section{Results and comparison with field data}
\label{ResultsDiscussion}
In this whole section, we discuss the results obtained with the
\emph{sugar lump} model and systematically compare with what we called
the \emph{standard} model
~\cite{cozar2014plastic,ter2016understanding,Eriksen2014}, that is to
say the case where fragments always break into two (identical) pieces
at each generation, whatever their size.  \par
\begin{figure*}[!ht]
\centering \includegraphics[width=0.7\textwidth]{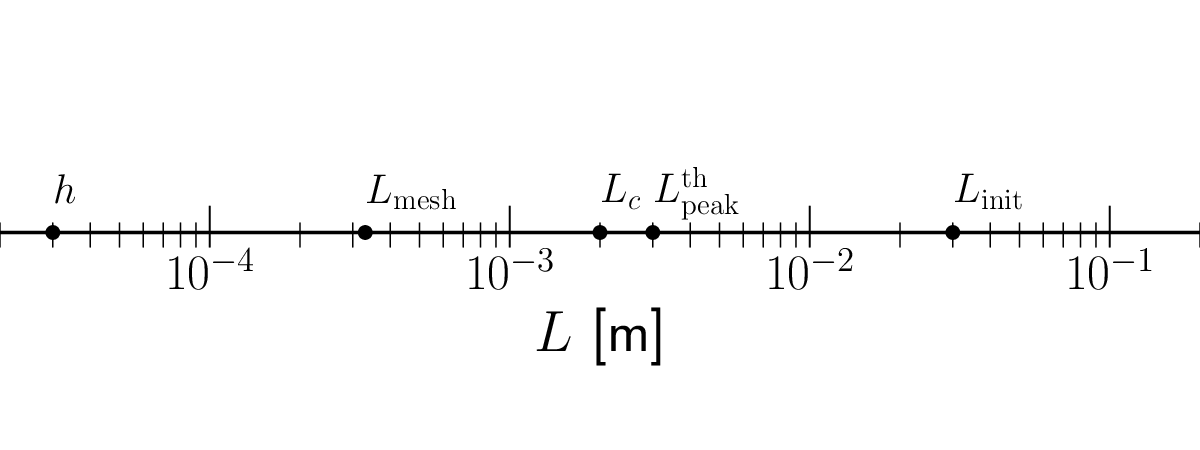}
\caption{Characteristic length scales: film thickness $h$, mesh size
  $L_{\mathrm{mesh}}$, breaking threshold $L_c$, predicted peak
  position in the size distribution of fragments
  $L_{\mathrm{peak}}^{\mathrm{th}}$, and largest lateral initial
  dimension of fragments $L_{\mathrm{init}}$. Values $h=30~\mu$m and
  $L_{\mathrm{init}}=3$~cm have been chosen for being typical
  values. A mesh size $L_{\mathrm{mesh}}=330$~$\mu$m is representative
  of neuston or manta nets used for collecting plastics debris. The
  peak position $L_{\mathrm{peak}}^{\mathrm{th}}$ actually depends on
  time, and may become smaller than the threshold length $L_c$ as
  fragmentation proceeds.}
\label{fig:scales}
\end{figure*}

 Whenever possible and
meaningful, we also compare our results with available field
data. Therefore, one needs to assign a numerical correspondence
between the physical time scale and the duration of a step in the
iterative models. The fragmentation rate of plastic pieces can be
assessed using accelerated aging
experiments~\cite{julienne2019influence,Julienne2022,menzel2022}. The
half-life time, corresponding to the time when the average particle
size is divided by 2, has been found around 1000~hours, which roughly
corresponds to one year of solar exposition~\cite{menzel2022}. Hence,
the iterative step $t$ used in all following sections can be
considered to be in the order of one year. For typical plastic film
dimensions, it is reasonable to assume that the thickness $h$ is
between 10 and 50~$\mu$m, and the initial largest lateral dimension
$L_{\mathrm{init}}$ is in the range of 1 to 5~cm. These characteristic
lengths, together with the other length scales involved in this paper
are positioned relative to each other in Fig.~\ref{fig:scales}.
\subsection{Evolution of the size distribution and of the total abundance of fragments with time}
\label{size distribution}
\begin{figure*}[!ht]
\centering
\includegraphics[width=0.7\textwidth]{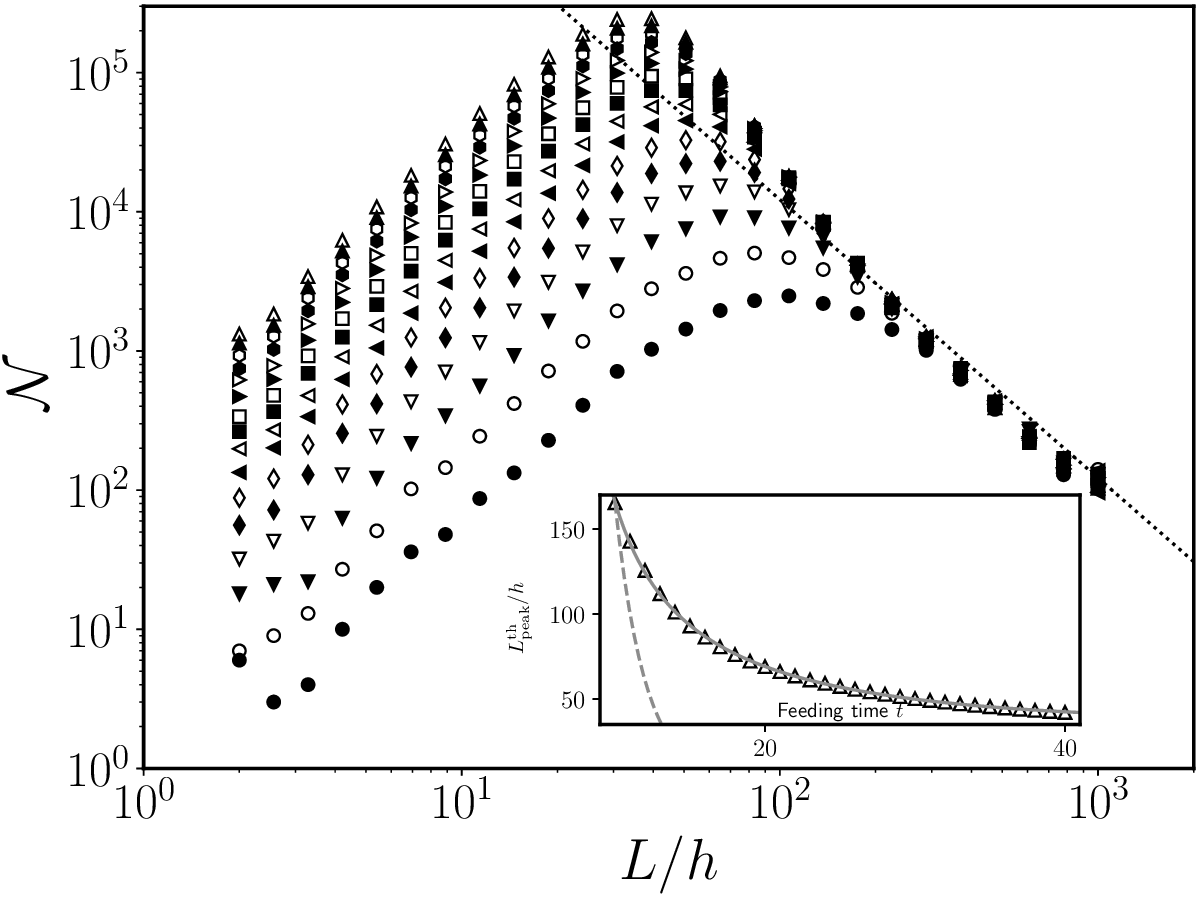}
\caption{Evolution of the size distribution of fragments over time for
  a chosen threshold value $L_c/h=100$. For clarity, the number of
  fragments $\cal N$ as a function of fragment size $L/h$ is shown
  from $t=10$ ($\bullet$) to $t=40$ ($\triangle$), every two time
  steps. The dotted line represents--in the \emph{standard} model--the
  expected size distribution.  Inset: Variation of the peak position
  $L_{\mathrm{peak}}^{\mathrm{th}}$ with time, the continuous line
  being merely a guide to the eyes. The dashed line is the
  exponentially-decaying evolution of the smallest size reached at a
  given time $t$ in the \emph{standard} model.}
\label{fig:SizeDistributionOverTime}
\end{figure*}

The size distribution of plastic fragments over time is represented in
Fig.~\ref{fig:SizeDistributionOverTime} for the \emph{sugar lump} and
confronted to the \emph{standard} model size distribution. The origin
of time corresponds to the date when the very first plastic waste was
dumped into the ocean.
\par According to the \emph{standard} model (see
Eq.~(\ref{eq:fillingLawAtomic2}), Section~\ref{standardModel}), the
amount of particles as a function of their size follows a power law of
exponent $-2$ which leads to a divergence of the number of particles
at very small size (dotted line in Fig
~\ref{fig:SizeDistributionOverTime}).  For large MP, the prediction of
the \emph{sugar lump} model was broadly similar, \emph{i.e.} following
the same power law.  By contrast, the existence of a mechanism
inhibiting the break of smaller objects, as introduced in the
\emph{sugar lump} model, did lead to the progressive built of an
abundance peak for intermediate size fragments due to the accumulation
of fragments with size around $L_c$ (see Section~\ref{sugarLump} for
details).  Moreover, the particle abundance at the peak increased with
time while the peak position shifted towards smaller size
classes. This shift was fast for the first generations, and then
slowed down when time passes: The inset in
Fig.~\ref{fig:SizeDistributionOverTime} showed how the existence of a
breaking threshold significantly slowed down the production of very
small particles compared to the standard model.  As can be observed
from the inset in Fig.~\ref{fig:SizeDistributionOverTime}, the peak
position $L_{\mathrm{peak}}^{\mathrm{th}}$, around $L_c$, decreased in
a small range typically between 1.5$L_c$ and 0.5$L_c$ for time periods
up to a few tens of years.

 %As mentioned in the introductory section (cf. section ~\ref{intro}) and explained in detail in a previous article ~\cite{george2021}, there are two families of MP in the marine environment, generated by two distinct degradation mechanisms: larger fragments generated by iterative bulk fragmentation of macro-waste, and smaller fragments obtained directly by delamination of thin surface layers of these macro-waste. These two families can be distinguished in Fig.~\ref{fig:resp2ref}, redrawn\footnote{To avoid the large dispersion arising from some empty size classes, the data has been smoothed over wider size classes--100~$\mu$m instead of 10~$\mu$m when required, that is to say for the large size end of the small MP size distribution. In Fig.~\ref{fig:resp2ref}, black triangles are associated to such reprocessed data} from data published in Ref.~\cite{Poulain2019}, which are, to the best of our knowledge, the only data allowing to compare directly the amount of MP over the whole range of size scales from $10$ to $10^4~\mu$m.

%
%\begin{figure}[!ht]
%\centering
%\includegraphics[width=0.7\columnwidth]{resp2ref.eps}
%\caption{Size distribution for small and large MP--(small black circles or black triangles, and large red circles, respectively--, redrawn from Ref.~\cite{Poulain2019}). Dotted black straight line and continuous red line: power laws $\propto1/L^2$ corresponding respectively to small and large MP families}
%\label{fig:resp2ref}
%\end{figure}

%In this paper, only bulk fragmentation is discussed. 
\par
\begin{figure*}[!ht]
\centering \includegraphics[width=0.7\textwidth]{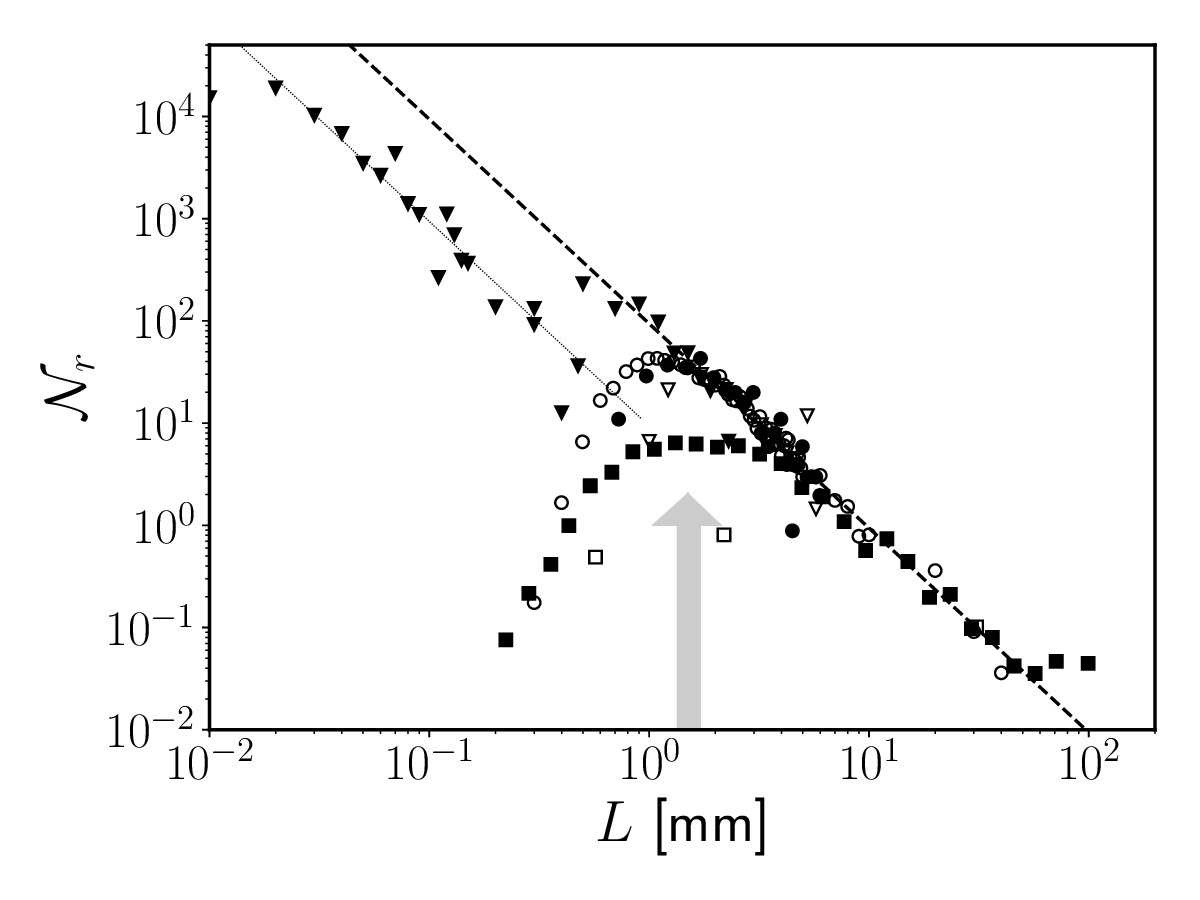}
\caption{Re-scaled size distribution of fragments ${\cal N}_r$ as a
  function of their largest dimension $L$ from various field data: $\blacktriangledown$--Poulain, Mercier \emph{et al.}~\cite{Poulain2019}; $\square$--Eriksen,
  Lebreton \emph{et al.}~\cite{Eriksen2014};
  $\blacksquare$--Coz\'ar, Etchevarr\'\i a \emph{et
  al.}~\cite{cozar2014plastic}; $\triangledown$--ter Halle, Ladirat
  \emph{et al.}~\cite{ter2016understanding}; $\bullet$--Tokai, Uchida
  \emph{et al.}~\cite{Tokai2021}; $\circ$--Isobe, Uchida \emph{et
  al.}~\cite{Isobe2015}. The two straight lines, dotted and dashed, are guides for the eye describing a $1/L^2$ scaling law.}
\label{fig:ceremonieusePlus}
\end{figure*}

 Let us now discuss the comparison to the experimental data. A
sample of various field data from different authors
~\cite{cozar2014plastic,ter2016understanding,Eriksen2014,Tokai2021,Isobe2015,
  Poulain2019} is displayed in Fig.~\ref{fig:ceremonieusePlus}.  In
order to obtain a collapse of the data points for large MPs, a
vertical scaling factor has been applied, since abundance values from
different sources can not be directly compared in absolute units. The
two main features of these curves are: A maximum abundance at a value
of a few millimeters (indicated by a gray arrow), and the collapse of
the data points onto a single $1/L^2$ master curve (indicated by a
dashed line). Note that data from Ref.~\cite{Poulain2019} (black
triangles) which is, to the best of our knowledge, the only data
allowing a direct comparison of the amount of MP over the whole range
of sizes from $10$ to $10^4~\mu$m, also demonstrates these features:
After a small abundance peak followed by a decrease, one observes
another increase when going to smaller sizes, as described in the
introduction. This high population of small MP is most probably the
result of delamination processes, which are not addressed here.

The threshold value $L_c$ is presumably defined by the energy balance
between the bending energy required for breaking a film and the
available turbulent energy of the ocean. The bending energy depends on
the film geometry and on the mechanical properties of the weathered
polymer. As shown for a fiber by Brouzet \emph{et
al}~\cite{BrouzetLongueurEl,Brouzet2021}, the threshold $L_c$ is
proportional to the fiber diameter $d$ and varies as
\begin{equation}
    L_c= k\frac{E^{1/4}}{(\rho\eta\epsilon)^{1/8}}d
    \label{eq:BrouzetLp}
\end{equation}
where $E$ is the Young modulus of the brittle polymer fiber, $\rho$
and $\eta$ are the mass density and viscosity of water, $\epsilon$ is
the mean turbulent dissipation rate and $k$ is a prefactor in the
order of 1.  In two dimensions, the expression for the threshold $L_c$
is more complex, since it depends both on the width $\ell$ and
thickness $h$ of the film. However, based on 2D mechanics, one can
show that the order of magnitude and $h$-dependency for $L_c$ remain
the same as in 1D, while the prefactor slightly varies with
$\ell$. Reasonable assumptions on film geometry, mechanical properties
of weathered brittle plastic and highly turbulent ocean events, such
as made by Brouzet \emph{et al.}~\cite{Brouzet2021} allowed us to
evaluate that $L_c/h \approx 100$.  For films of typical thicknesses
lying between 10 and 50~$\mu$m, this gave a position of the peak
between 1 and 5~mm in good agreement with the field data represented
in Fig.~\ref{fig:ceremonieusePlus}. It would be delicate to refine
more since plastic waste in the environment exhibits a large range of
mechanical properties due to their various chemical nature (including
oxidation state), macromolecular architecture, the presence of
inorganic fillers, etc.

\par It is also interesting to
discuss the power law exponent value exhibited by both \emph{standard}
and \emph{sugar-lump} models at large MP sizes. In time-invariant
models, the theoretical exponent actually varies with the
dimensionality of the considered objects (fibers, films, lumps)
ranging from $-1$ (fibers) to $-3$ (lumps).  As expected, when the
objects dimensionality is fixed, the value $-2$ observed in
Fig.~\ref{fig:SizeDistributionOverTime} for the \emph{sugar-lump}
model is due to the hypothesis of film-like pieces breaking along
their larger dimension only, keeping their thickness constant. In the
same way, regarding the laboratory experiments performed on glass
fibers~\cite{Brouzet2021}, the large MP distribution is compatible in
the long-time limit with the expected $-1$ power
law~\footnote{provided that, of course, the depletion of very large
objects that originates from the absence of feeding is disregarded.}.
Coming back to the field data as displayed in
Fig.~\ref{fig:ceremonieusePlus}, one can note that for large MP all
data points collapsed onto a single $1/L^2$ master curve. This suggests
that either most collected waste comprises film-like objects breaking
along their larger dimension only, or, perhaps more likely, that one
collects a mixture of all three types of objects leading to an
``average'' exponent, obviously lying somewhere between $-1$ and $-3$,
that turns out to be close to $-2$.  \par 

\par
\begin{figure*}[!ht]
\centering
\includegraphics[width=0.7\textwidth]{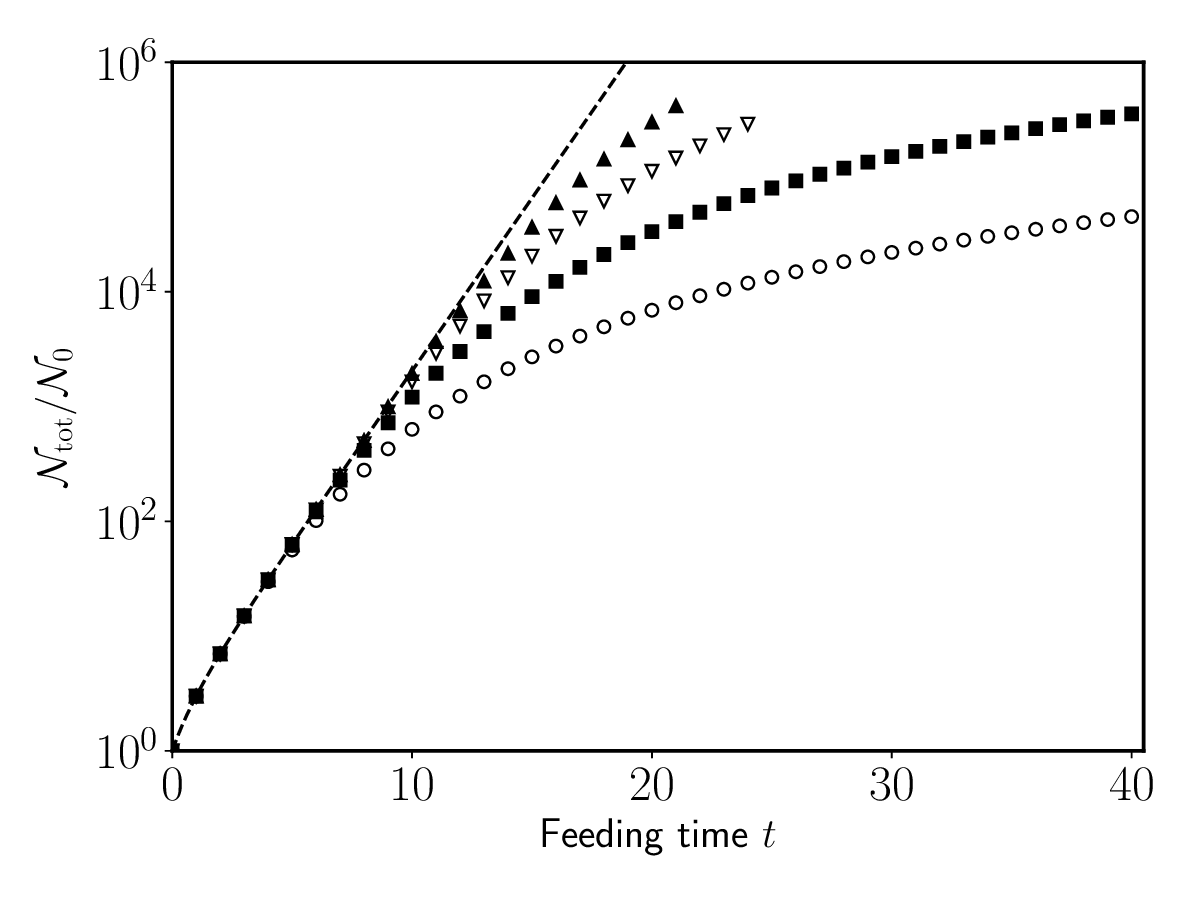}
\caption{Total amount of fragments in the ocean (reduced units) as a
  function of time $t$ (\emph{i.e.} number of iterative steps. One
  step approximately corresponds to one year), for different values of
  the size threshold $L_c$: $L_c=100h$--$\circ$;
  $L_c=30h$--$\blacksquare$; $L_c=10h$--$\triangledown$;
  $L_c=3h$--$\blacktriangle$. Dashed line: \emph{standard} model. A
  value ${\cal N}_0=30$ has been chosen here}
\label{fig:abundanceTot}
\end{figure*}

The total abundance ${\cal
  N}_{\mathrm{tot}}$ of fragments (all sizes included) as a function
of time was represented in Fig.~\ref{fig:abundanceTot} for both the
\emph{sugar lump} and \emph{standard} models.  In the latter case, the
abundance was simply described by an exponential law: ${\cal
  N}_{\mathrm{tot}} = [2^{t+1}-1]{\cal N}_0$ when the ocean was fed by
a constant number ${\cal N}_0$ of (nearly identical) large fragments
per iteration (Eq.~\ref{eq:cumAbundanceStandard},
Section~\ref{standardModel}). The \emph{sugar lump} model predicted a
time evolution which deviates from the \emph{standard} model
prediction: The increase of total abundance slowed down with time, due
to the hindering of smaller fragments production, and the effect was
all the more pronounced for larger threshold parameters $L_c$, as
could have been expected.  In the realistic case where $L_c/h \approx
100$, the increasing rate of fragments production became very small
for the largest feeding times, as can be observed in
Fig.~\ref{fig:abundanceTot} which showed that the number of MP would be
multiplied every ten years by only a factor 2, compared to a factor of
1000 in the standard model. These theoretical results might explain
why no clear temporal trend is observed in the field
data~\cite{galgani2021}.

\subsection{Role of the mesh size on the size distribution and on its temporal evolution}
\label{transitory and steady-state}
If one wants to go further in confronting models to field data, one
needs to take into account that the experimental collection of
particles in the environment always involves an observation window,
and in particular a lower size limit $L_{\mathrm{mesh}}$, \emph{e.g.}
due to the mesh size of the net used during ocean campaigns. The very
existence of a lower limit leads to the appearance of transitory and
steady-state regimes for the temporal evolution of the number of
collected particles, as will be shown below.  \par
\begin{figure*}[!ht]
\centering
\includegraphics[width=0.7\textwidth]{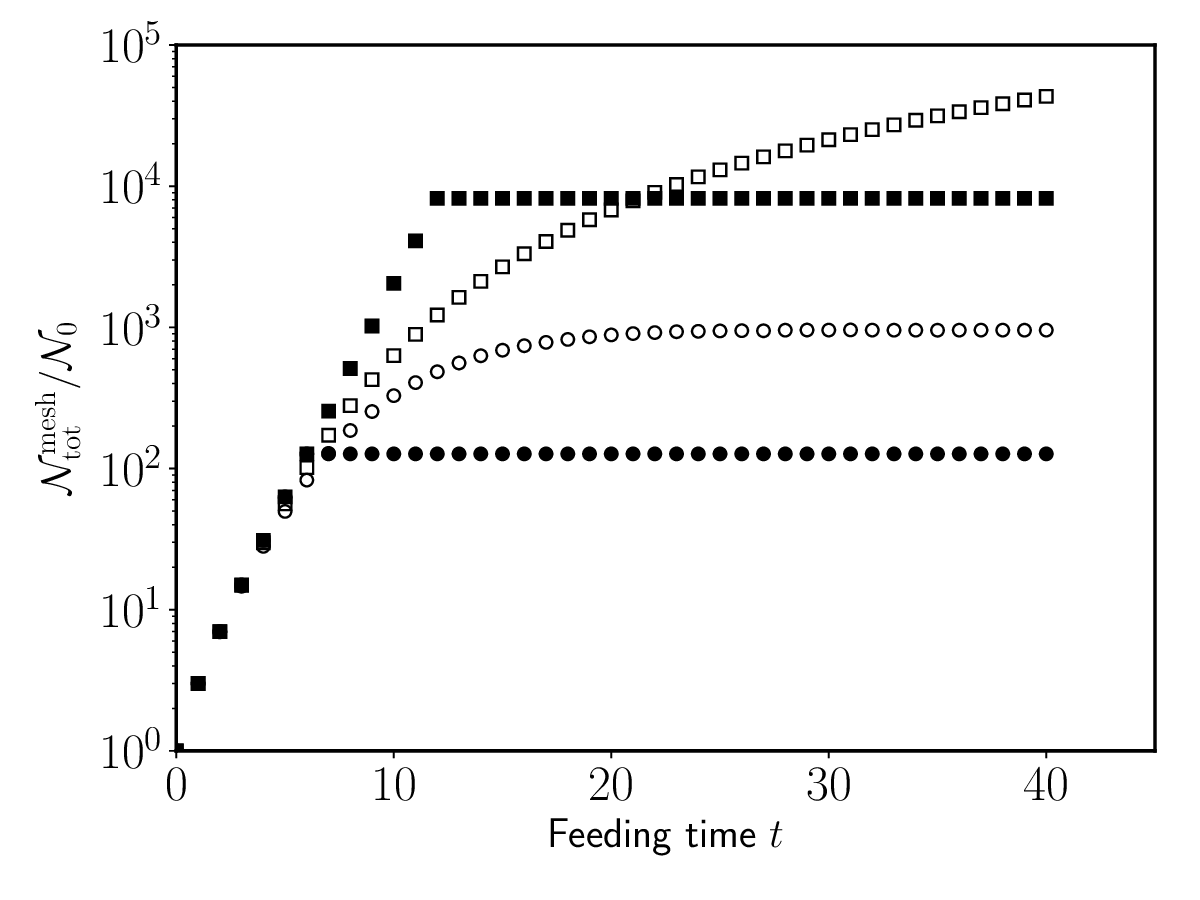}
\caption{Time dependence of the total number of collected fragments
  ${\cal N}_{\mathrm{mesh}}$ (reduced units) for two different values
  of the mesh size in the case of the \emph{standard} model (full
  symbols) and of the \emph{sugar lump} model (empty symbols) with
  threshold $L_c/h=100$. Mesh size values for different symbol keys
  are, for the \emph{standard} model, $L_{\mathrm{mesh}}/h$:
  $\bullet$, 93; $\blacksquare$, 6 and for the \emph{sugar lump} model
  $L_{\mathrm{mesh}}/h$: $\circ$, 92; $\square$, 7}
\label{fig:sievedStandardSugarLump-LARGE_Lmesh-threshold100}
\end{figure*}

 In the
\emph{standard} model case, when the feeding and breaking process
started, larger size classes were first filled, while smaller size
classes were still empty (Fig.~\ref{fig:remplissagecase2},
Section~\ref{standardModel}).  As long as the smaller fragments
produced by the breaking process were larger than the lower size limit
$L_{\mathrm{mesh}}$ of the collection tool, the number of collected
fragments increased with time, \emph{de facto} producing a transitory
regime in the observed total abundance. The size of the smaller
fragments reached $L_{\mathrm{mesh}}$ after a given number of
fragmentation steps corresponding to the duration of the transitory
regime:
\begin{equation}
    t_c\approx2\ln(L_{\mathrm{init}}/L_{\mathrm{mesh}})/\ln2
\label{eq:Tempstransitoire}
\end{equation}
where $L_{\mathrm{init}}$ is the initial largest dimension of the
plastic fragments released into the ocean.  From this time onward,
both the size distribution and total number of collected fragments in
the observation window no longer changed. Even though the production
of fragments smaller than $L_{\mathrm{mesh}}$ continued to occur, as
well as the continuous feeding of large-scale objects, one therefore
observed a steady-state regime. This was illustrated in
Fig.~\ref{fig:sievedStandardSugarLump-LARGE_Lmesh-threshold100} for
two different values of the mesh size $L_{\mathrm{mesh}}$ (full
symbols $\bullet$ and $\blacksquare$).

\par For the \emph{sugar lump} model case, one needed to also consider
the size threshold length scale $L_c$, below which fragmentation is
inhibited. When $L_c$ was much smaller than $L_{\mathrm{mesh}}$, the
threshold length $L_c$ was not in the observation window, hence the
analysis was the same as in the \emph{standard} case. At contrast,
when $L_c$ was close to $L_{\mathrm{mesh}}$ or larger, the transitory
regime was expected to exhibit two successive time dependencies. This
behavior was displayed in
Fig.~\ref{fig:sievedStandardSugarLump-LARGE_Lmesh-threshold100} (empty
symbols $\circ$ and $\square$) for the same mesh size values as in the
\emph{standard} model for comparison. At short times, since the
smaller fragment size had not reached yet the breaking threshold
$L_c$, the number of collected fragments followed the same law as in
the \emph{standard} case. When the smaller fragments got close to the
size $L_c$, however, the inhibition of their breaking created an
accumulation of fragments around $L_c$, hence the abundance peak. As a
consequence, the increase in the total number of fragments slowed
down. Since the abundance peak position shifted towards smaller values
with time (Fig.~\ref{fig:SizeDistributionOverTime}, inset) albeit
slowly, a final stationary state should be observed when the abundance
peak position became significantly smaller than
$L_{\mathrm{mesh}}$. As shown in
Fig.~\ref{fig:sievedStandardSugarLump-LARGE_Lmesh-threshold100}, this
occurred within the explored time window for large $L_{\mathrm{mesh}}$
($\circ$), but the stationary state was not observed for small
$L_{\mathrm{mesh}}$ ($\square$), presumably because our simulation had
not explored times large enough.  When the steady-state regime was
reached, the number of fragments above $L_{\mathrm{mesh}}$,
\emph{i.e.} likely to be collected, remained constant with a value
larger than that of the standard model, due to the overshoot induced
by the accumulation on the right-hand side of the peak.

\par Let us recall that the characteristic fragmentation time, defined
as the typical duration for a piece to break into two, has been
evaluated at one year. In the case of the \emph{standard} model, this
meant that the size of each fragment was reduced by a factor 30 in
about 10 years. Therefore, starting with debris size of the order of a
centimeter, small MPs of typical size the mesh size (330~$\mu$m in
Fig.~\ref{fig:scales}) would be obtained within 10 years only. Thus, 10
years correspond to the duration of the transitory regime $t_c$
established in Eq.~(\ref{eq:Tempstransitoire}) and the oceans should
be by far in the steady-state regime since the pollution started in
the 1950's. It is however no longer controversial nowadays that the
\emph{standard} (steady-state) model fails to describe the size
distribution of the field data.  On the contrary, the \emph{sugar
lump} model predicted the existence of an abundance peak, in agreement
with what was observed during collection campaigns. This peak is due to
the accumulation of fragments whose size is in the order of the
breaking threshold $L_c$.  As discussed in subsection \ref{size
  distribution}, the failure threshold $L_c$ could be soundly estimated
to lie between $1$ and $5$ mm. Comparison with field data then
corresponded to the case where $L_c$ was about ten times larger than the
mesh size $L_{\mathrm{mesh}}$. As just shown in
Fig.~\ref{fig:sievedStandardSugarLump-LARGE_Lmesh-threshold100}, this
implied a drastic increase of duration of the transitory regime, that
could be estimated to be above 100 years. These considerations lead us
to the important conclusion that one is still nowadays in the
transitory regime. Moreover, the \emph{sugar lump} model also implies
that the total abundance is correctly estimated through field data
collection, \emph{i.e.} that it is not biased by the mesh
size. Because the peak position slowly shifted towards smaller sizes,
the mesh size would eventually play a role, but at some much later
point in time.  Finally, let us recall that this paper did not take
into account delamination processes, so the previous statement is only
true for millimetric debris, that is to say debris produced through
fragmentation, and that micrometric size debris might exhibit a
completely different behavior, being in addition much more numerous.

\subsection{Constant versus exponential feeding}
\label{ExpGrowth}
\begin{figure}[!ht]
\centering \includegraphics[width=0.7\columnwidth]{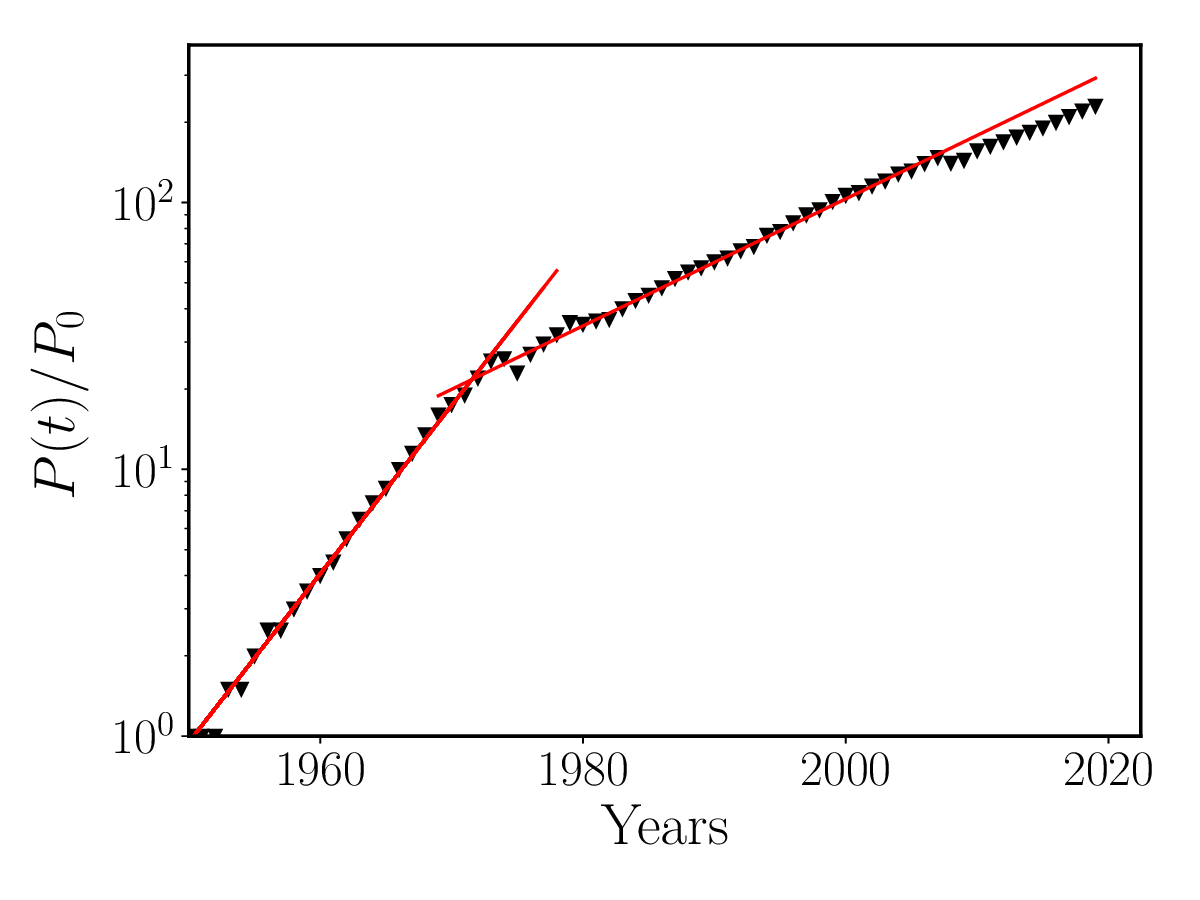}
\caption{Global plastics production per year from 1950 onward,
  relative to $P_0=2\times10^6$~t produced in 1950. A significant
  break in plastics production is observed in 1974. Solid lines:
  annual growth rates, about 16\% or 6\% per year before or after
  1974, respectively}
\label{fig:cochon}
\end{figure}

\begin{figure}[!ht]
\centering \includegraphics[width=0.7\columnwidth]{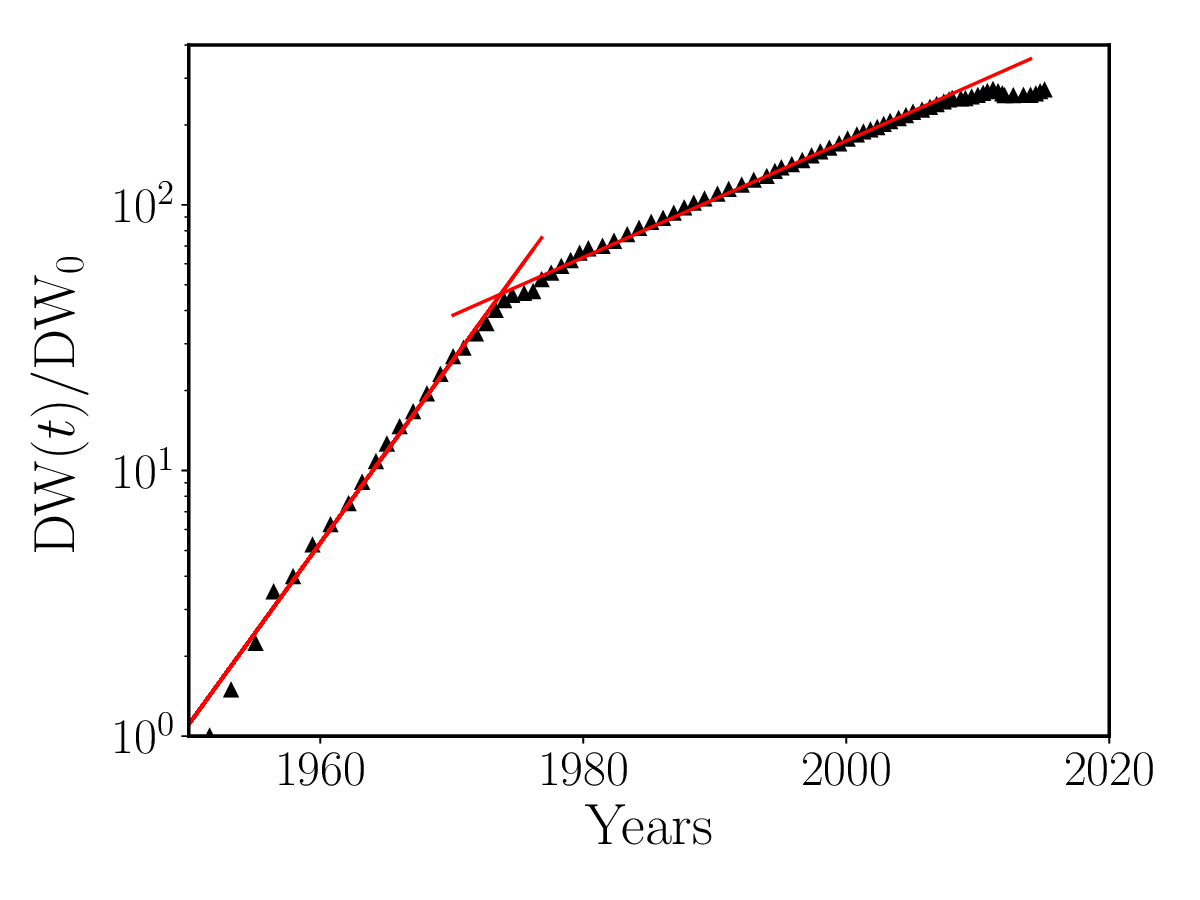}
\caption{Estimated discarded plastic wastes, in relative units. A
  significant break in discarded waste is observed in 1974. Solid
  lines: annual growth rates, about 17\% or 5\% per year before or
  after 1974, respectively}
\label{fig:jetCochon}
\end{figure}

In the results discussed in Section~\ref{size distribution}, it was
assumed that the rate of waste feeding in the ocean is constant with
time. However, it is common knowledge that the production of plastics
has increased significantly since the 1950's. Geyer \emph{et
al}~\cite{Geyer2017} show that the discarded waste follows the
same trend. Data from the above-quoted article has been extracted and
fitted in Fig.~\ref{fig:cochon} and Fig.~\ref{fig:jetCochon} with
exponential laws ${\cal N}={\cal N}_0(1+\tau)^t$, where $\tau$
represents an annual growth rate, of plastic production and discarded
waste respectively. For plastic production, the annual growth rate is
found about 16$\%$ until 1974, the year of the oil crisis, and close
to 6$\%$ after 1974 with, perhaps, an even further decrease of the
rate in the recent years. Not unexpectedly, the same trends are found
when considering the \emph{discarded} waste, with growth rates,
respectively, 17\% and 5\%.  In order to discuss now the effects of an
increasing waste feeding in the ocean, we injected for simplicity a
single exponential with an intermediate rate of 7$\%$ in the two
models.

\begin{figure*}[!ht]
\centering \subfloat[][14 iteration
  steps]{\includegraphics[scale=1,width=0.7\textwidth]{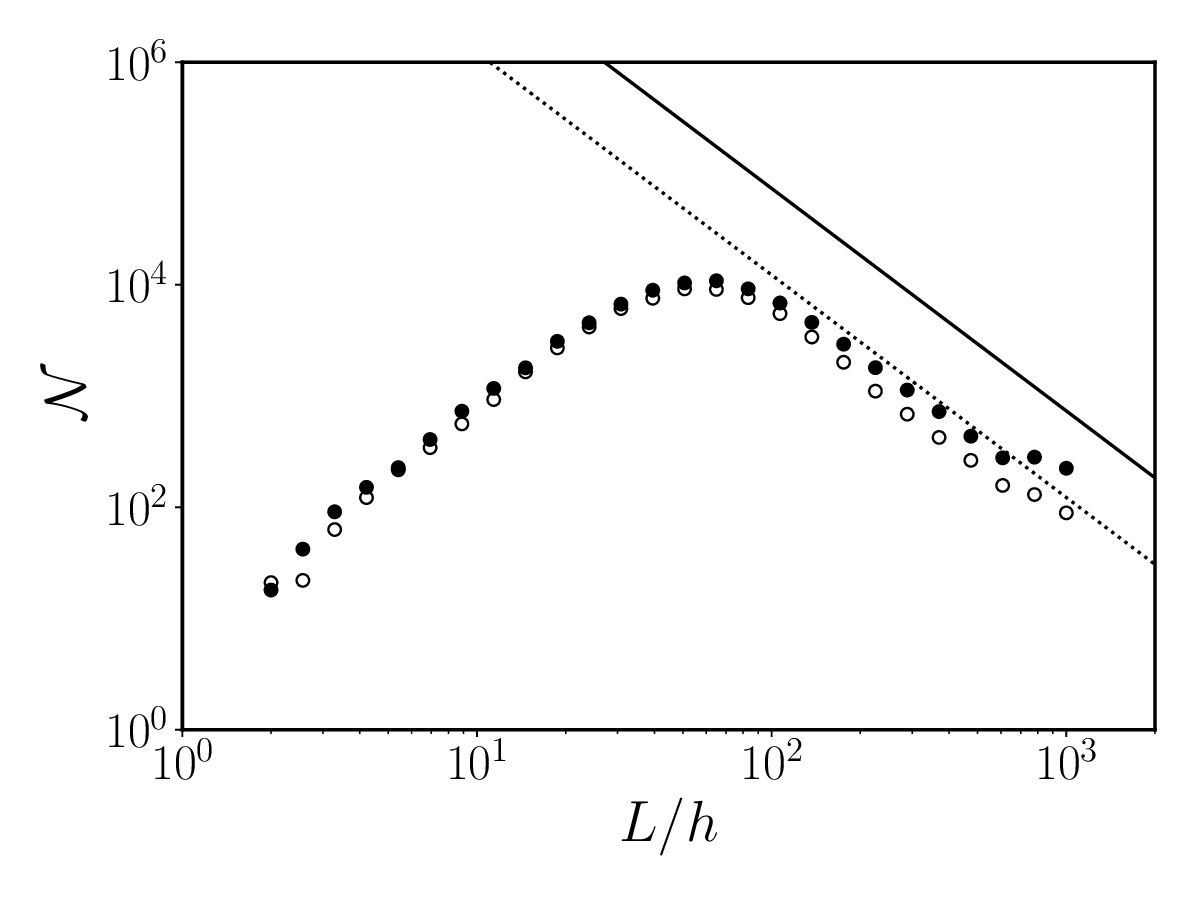}\label{fig:sizeDistributionAvecSansInflation-14-threshold100}}\\ \subfloat[][40
  iteration
  steps]{\includegraphics[scale=1,width=0.7\textwidth]{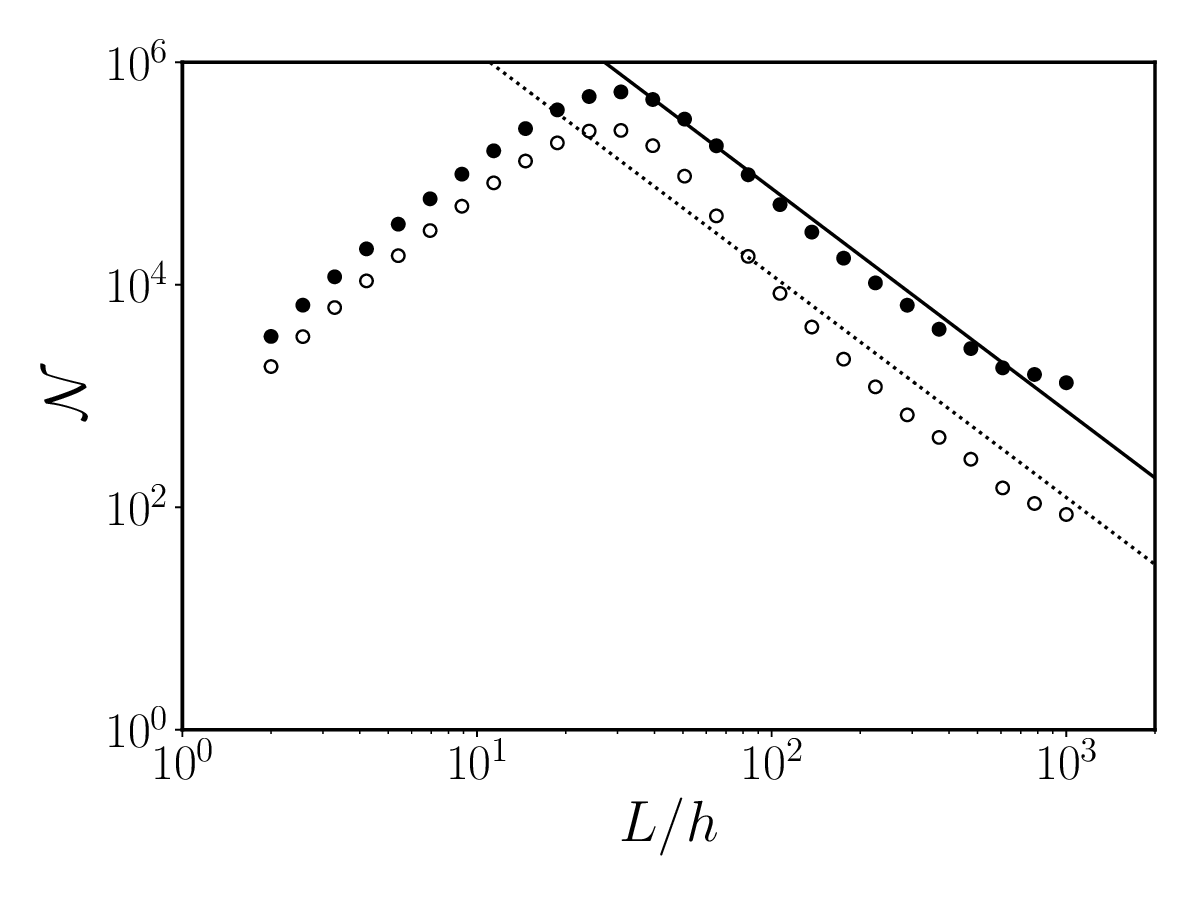}\label{fig:sizeDistributionAvecSansInflation-40-threshold100}}
\caption{Comparison of the size distributions of fragments $\cal N$
  for the two feeding modes : $\bullet$ exponential feeding, $\circ$
  constant feeding, for (a) a small number of fragmentation steps and
  (b) a large number of fragmentation steps. Continuous and dotted
  straight lines are guides for the eyes for the --$L^{-2}$ scaling
  law
\label{fig:sizeDistributionAvecSansInflation-threshold100}}
\end{figure*}

When comparing this feeding law and the \emph{standard} fragmentation
law $[2^{t+1}-1]{\cal N}_0$, one easily concluded that the total
number of plastics items in the ocean was mainly determined by the
fragmentation rate, regardless of the feeding rate. In order to
verify what happens in the case of the \emph{sugar lump} model, where
the fragmentation process was hindered, the size distributions for both
feeding hypotheses were numerically compared in
Figs.~\ref{fig:sizeDistributionAvecSansInflation-14-threshold100} and
\ref{fig:sizeDistributionAvecSansInflation-40-threshold100},
respectively after 14 and 40 years.
It could be observed that at short times, the size distribution was very
little altered by the change in feeding. At longer times, a
significant increase of the amount of the largest particles could be
observed, while the amount of small particles was increasing much
less. Besides, the size position of the abundance peak was almost not
shifted.
 
\begin{figure*}[!ht]
\centering
\includegraphics[width=0.7\textwidth]{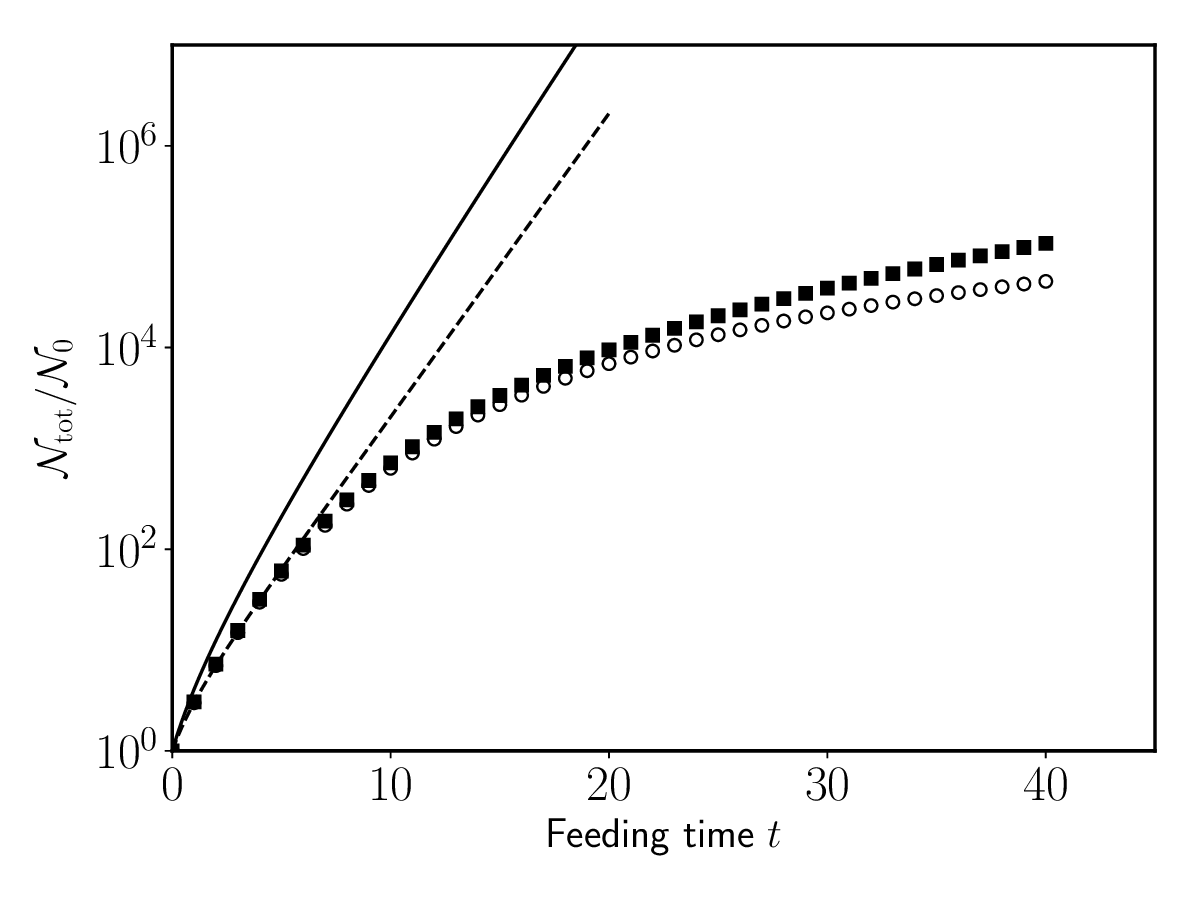}
\caption{Total amount of fragments in the ocean (reduced units) as a
  function of time $t$ (arbitrary units), for standard (lines) and
  \emph{sugar lump} (symbols) models and two feeding hypothesis,
  namely constant or exponential feedings. The inflation rate is 0
  (empty symbols $\circ$ or dashed line) or $7$\% (full symbols
  $\blacksquare$ or solid line).  For the \emph{sugar lump} model, the
  size threshold is $L_c=100h$}
\label{fig:abundanceTot-generation-avecSansInflation}
\end{figure*}

The total amount of fragments was represented in
Fig.~\ref{fig:abundanceTot-generation-avecSansInflation} for the
\emph{standard} and \emph{sugar lump} models for the two feeding cases
considered. For exponential feeding, the \emph{sugar lump} model still
predicted a significant decrease in the rate of fragment generation
over time, whereas one could have thought that exponential feeding
could cancel out this slowdown.
The conclusions drawn above, Section~\ref{transitory and steady-state}
therefore remained valid in the more realistic case of an exponential
feeding.

Finally, one should keep in mind that, if the feeding rate is a
reasonable indicator of plastic pollution, since it describes the
evolution over time of the \emph{total mass} of plastics present in
the ocean, it is not enough to properly describe plastic pollution.
For a given mass, the number--hence size--of particles produced is the
major factor in assessing potential impacts. Indeed, the smaller the
size the larger the particles number concentration, the larger their
specific area hence their adsorption ability and the larger the
ensuing eco-toxicity.  It was shown here that the mass of waste roughly
doubles every 10 years, whereas the number of particles doubles every
year, making fragmentation the main factor driving plastic pollution
and impacts.  A lot of studies are devoted to making a mass balance
and understanding the fluxes of plastic waste
~\cite{Geyer2017,Lebreton2019,Cordier2019,Sonke2022} but, even in the
case of a drastic immediate reduction of waste production, plastic
pollution and impacts will affect the ocean life for still many years
to come, due to fragmentation.

\section{Conclusion}
The generalist model presented here was based on a few sound physical
assumptions and shed new light on global temporal trends in the
distribution of microplastics at the surface of the oceans. The model
showed that the existence of a physical size threshold below which
fragmentation is strongly inhibited, lead to the accumulation of
fragments at a given size, in line with what is observed in the field
data. In other words, if one does collect much fewer particles in the
range $150$~$\mu$m--$1$~mm ~\cite{Poulain2019}, it is because only a
few of them is actually generated by fragmentation at this scale. One
would not necessarily need to invoke any other mechanism or bias such
as ingestion by living organisms~\cite{Dawson2018} or the mesh size of
collection nets~\cite{Isobe2019,Lindeque2020}, to explain the field
data for floating debris. As a consequence, the observed distribution
does reflect, in our opinion, the real distribution of MPs at the
surface of the ocean, down to 100~$\mu$m.  Besides, the \emph{sugar
lump} model implied a slowdown in the rate of MPs production by
fragmentation, due to the fact that fragmentation is inhibited when
particles approach the threshold size. This may explain the absence of
a clear increase in the MP numbers in different geographical
areas~\cite{galgani2021}.  \par Two other general facts have been
pointed out in this paper:
\begin{itemize}
    \item For large MP, the predicted size distribution followed a
      power law, whose exponent depends on the dimensionality of the
      object (-1 for a fiber, -2 for a film and -3 for a lump). It is
      therefore worth sorting out collected objects according to their
      geometry, as it is done for instance when fibers are separated
      from 2D objects~\cite{Lindeque2020}. It is however interesting
      to note that, when the objects are not sorted in this way, an
      ``average value'' -2 is found for the exponent.
    \item The model took into account an exponentially-increasing
      waste feeding rate. We have fitted the plastic production since
      the 1950's and found that there is not one but two exponential
      laws, the second one, slower than the first one, being visible
      after the oil crisis in 1974. Comparing this feeding to the
      exponential fragmentation ratio, we showed that the number of
      fragments was mainly predicted by the fragmentation process,
      regardless of the feeding details.
\end{itemize}
\par To go further and estimate absolute values of MP concentrations in the
whole range of sizes, it would be necessary, on one hand, to take into
account delamination in order to get small particles distribution. On
the other hand, one should also be aware of the spatial heterogeneity
of particles concentration and therefore an interesting development
could be to combine fragmentation with flow models developed for
instance in Refs.~\cite{Sonke2022,vanSebille2020}.
\section*{CRediT author statement}
\noindent \textbf{Matthieu \textsc{George}:~} Conceptualization,
Validation, Writing -- Review \verb+&+ Editing, Project
administration\\ \textbf{Frédéric \textsc{Nallet}:~} Methodology,
Software, Writing -- Review \verb+&+ Editing\\ \textbf{Pascale
  \textsc{Fabre}:~} Conceptualization, Validation, Writing -- Review
\verb+&+ Editing, Funding acquisition
\section*{Declaration of interests}
The authors declare that they have no known competing financial
interests or personal relationships that could have appeared to
influence the work reported in this paper
\appendix
\section{Supporting information}
\label{suppInfo}
\subsection{Standard model}
\label{standardModel}
\begin{figure}[!ht]
\centering \includegraphics[width=0.7\columnwidth]{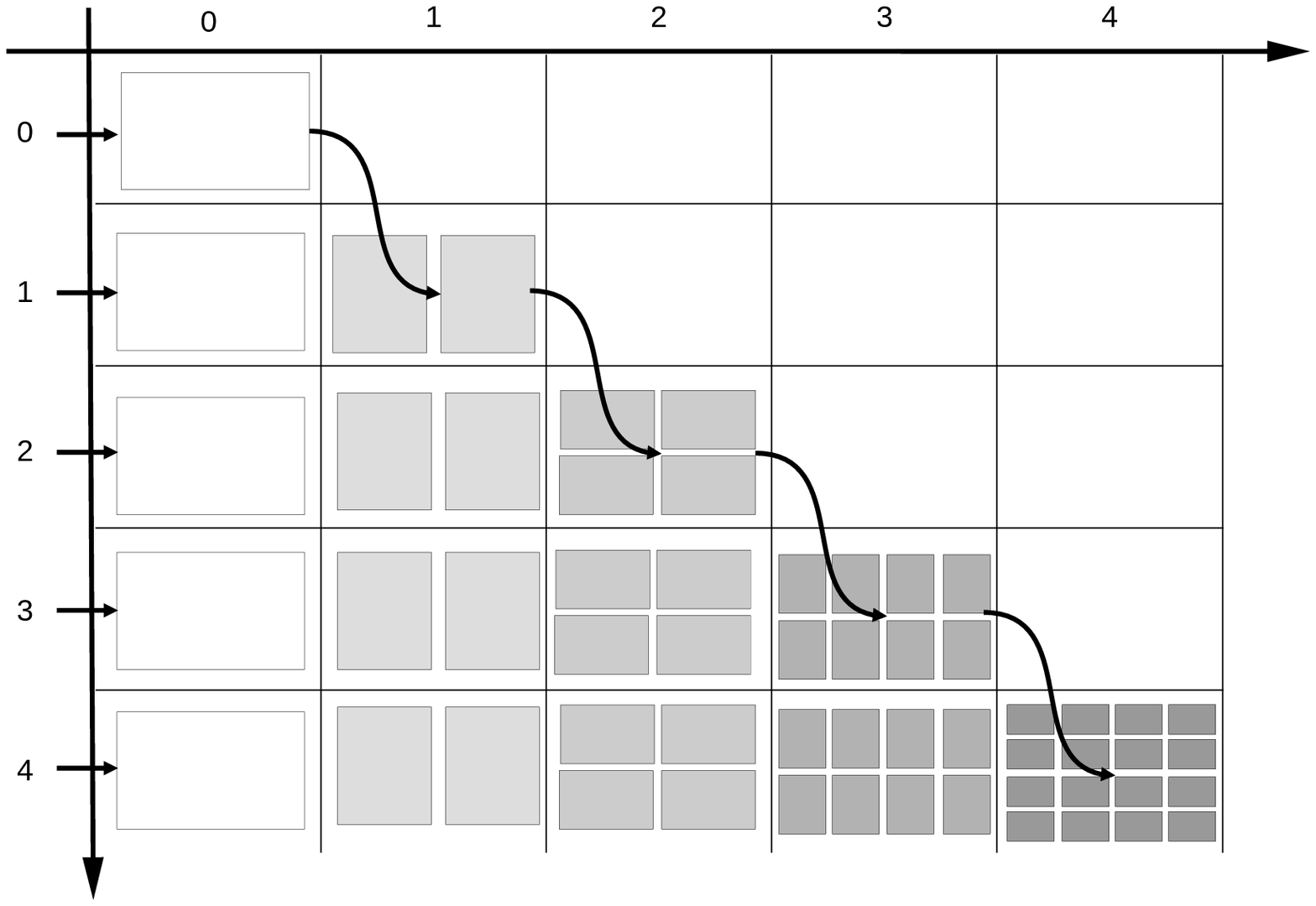}
\caption{Schematics of iterative size classes filling by the standard
  model. Horizontal axis, rightwards: size (or generation) index $p$;
  Vertical axis, downwards: time index $n$. Short horizontal arrows
  stand for the feeding process. Wavy arrows stand for the
  fragmentation process. The shade of gray represents the
  \emph{weathering} of a given object, that is to say the time spent
  in the ocean}
\label{fig:remplissagecase2}
\end{figure}

In this model, as pictorially represented in
Fig.~\ref{fig:remplissagecase2}, the ocean is fed at each iteration
$n$ with a fixed number $a_0$ of large 2D-like objects, mimicking
plastic films.
Neglecting size and shape dispersity for convenience, all
0$^{th}$-generation objects are assumed to be large square platelets
of lateral size $L_{\mathrm{init}}$ and thickness $h$, with
$L_{\mathrm{init}}\gg h$. Between consecutive iteration steps,
fragmentation produces $p^{th}$-generation objects, by splitting in
two equal parts $(p-1)^{th}$-generation objects, thus generating
\emph{square} platelets when $p$ is even, but \emph{rectangular}
platelets with aspect ratio 2:1 for odd $p$. If size is measured by
the diagonal, a $p^{th}$-generation object has size
$\sqrt{2}L_{\mathrm{init}}/2^{p/2}$ (even $p$) or
$\sqrt{5}L_{\mathrm{init}}/2^{(p+1)/2}$ (odd $p$). With size classes
described by the number of $p^{th}$-generation objects at iteration
step $n$, $C(n,p)$, the filling law of size classes is:
\begin{equation}
\begin{array}{rcll}
C(n,0)&=&a_0& \\ C(n,p)&=&0&\mathrm{~if~}p>n
\\ C(n,p)&=&2C(n-1,p-1)&\mathrm{~if~}1\leq p\leq n
\end{array}
\label{eq:fillingLawStandard}
\end{equation}
The set of equations~(\ref{eq:fillingLawStandard}) is readily solved:
$C(n,p)=2^pa_0$ for $0\leq p\leq n$, and $C(n,p)=0$ for $p>n$. Since
size $L$ scales with generation index $p$ as $2^{-p/2}$, the
steady-sate scaling for the filling of size classes is $C\propto
L^{-2}$.
\par The cumulative abundance $S_n\equiv\sum_pC(n,p)$ at iteration
step $n$ is also easily obtained:
\begin{equation}
    S_n=\left[2^{n+1}-1\right]a_0
    \label{eq:cumAbundanceStandard}
\end{equation}
and displayed as a dashed line in Figs.~\ref{fig:abundanceTot} and
\ref{fig:abundanceTot-generation-avecSansInflation}.
\par
\begin{figure}[!ht]
\centering \includegraphics[width=0.7\columnwidth]{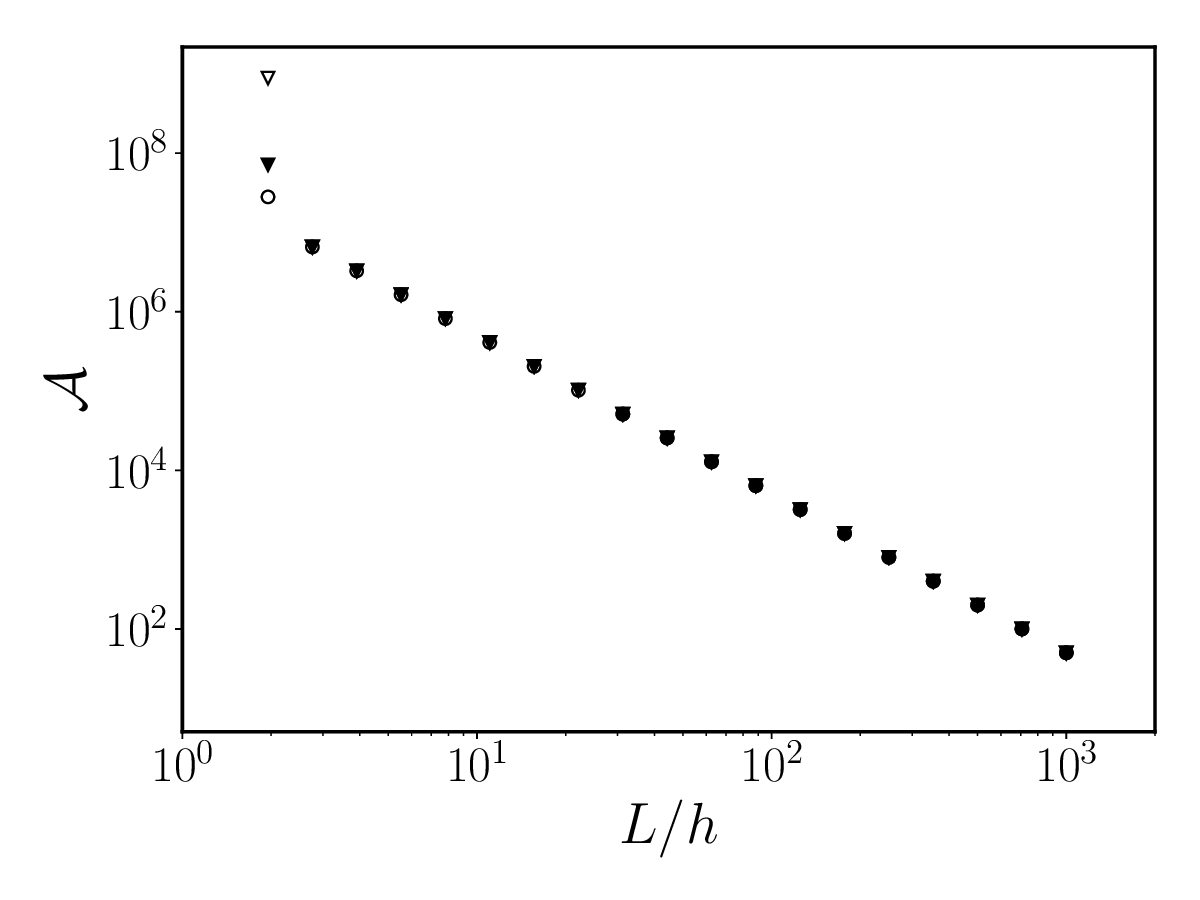}
\caption{Time-invariant features of the standard model, as long as the
  ``atomic limit'' has not been reached yet. Symbol keys and time
  indices: $\bullet$, $n=10$; $\circ$, $n=20$; $\blacktriangledown$,
  $n=40$; $\triangledown$, $n=80$. ``Atomic'' limit reached for
  generation index $p_{\mathrm{max}}$ chosen equal to $18$}
\label{fig:atomicLimit}
\end{figure}

 As noticed in Ref.~\cite{ter2016understanding} where experimental
data and model predictions are matched together, the standard model
fails for small objects, and this occurs when a (nearly) cubic shape
is reached. Since the typical (lateral) size of $p^{th}$-generation
objects is $\approx L_{\mathrm{init}}/2^{p/2}$, the limit is reached
for
\begin{equation}
    p_{\mathrm{max}}\approx2\frac{\log\frac{L_{\mathrm{init}}}{h}}{\log2}
    \label{eq:standardFailure}
\end{equation}
that is to say in about 20 generations with the rough estimate
$L_{\mathrm{init}}/h=10^3$. The set of equations describing the
size-class filling law has to be altered to take into account this
limit. Assuming for simplicity that $p_{\mathrm{max}}$-generation
objects cannot be fragmented anymore (``atomic'' fragments), this set
of equations becomes:
\begin{widetext}
\label{eq:fillingLawAtomic}
\begin{equation}
\begin{array}{rcll}
C(n,0)&=&a_0&
\\ C(n,p)&=&0&\mathrm{~if~}p>n\mathrm{~or~}p>p_{\mathrm{max}}
\\ C(n,p)&=&2C(n-1,p-1)&\mathrm{~if~}1\leq
p<p_{\mathrm{max}}\mathrm{~and~}p \leq n
\\ C(n,p_{\mathrm{max}})&=&C(n-1,p_{\mathrm{max}})+2C(n-1,p_{\mathrm{max}}-1)&\mathrm{~if~}n>p_{\mathrm{max}}
\end{array}
\end{equation}
\end{widetext}
As shown by the explicit solution, Eq.~(\ref{eq:fillingLawAtomic2})
below, the last line in this set of equations leads to an
\emph{accumulation} of ``atomic'' fragments (see also
Fig.~\ref{fig:atomicLimit} for a pictorial representation of this
feature)
\begin{widetext}
\begin{equation}
    \label{eq:fillingLawAtomic2}
    \begin{array}{rcll}
    C(n,p)&=&2^pa_0&\mathrm{~if~}0\leq p\leq n<p_{\mathrm{max}}
    \\ C(n,p_{\mathrm{max}})&=&\left(n+1-p_{\mathrm{max}}\right)2^{p_{\mathrm{max}}}a_0&\mathrm{~if~}n\geq
    p_{\mathrm{max}} \\ C(n,p)&=&0&\mathrm{~for~other~cases}
    \end{array}
\end{equation}
\end{widetext}
associated to a \emph{significant} (exponential to linear) slowing
down of the cumulative abundance:
\begin{equation}
    S_n=\left[2^{p_{\mathrm{max}}}\left(2+n-p_{\mathrm{max}}\right)-1\right]a_0
    \label{eq:cumAbundanceAtomic}
\end{equation}
for iteration steps $n\geq p_{\mathrm{max}}$.
\subsection{Standard model with inflation}
\label{inflationaryStandard}
As a first extension of the standard model, \emph{inflation} in the
feeding of the ocean with large 2D-like objects is now
considered. Taking simultaneously into account the ``atomic'' nature
of small fragments beyond $p_{\mathrm{max}}$ generations, the
size-class filling set of equations~(\ref{eq:fillingLawStandard}) has
to be replaced by:
\begin{widetext}
\begin{equation}
\label{eq:inflationaryStandard}
\begin{array}{rcll}
C(n,0)&=&a_0\left(1+\tau\right)^n&
\\ C(n,p)&=&0&\mathrm{~if~}p>n\mathrm{~or~}p>p_{\mathrm{max}}
\\ C(n,p)&=&2C(n-1,p-1)&\mathrm{~if~}1\leq
p<p_{\mathrm{max}}\mathrm{~and~}p \leq n
\\ C(n,p_{\mathrm{max}})&=&C(n-1,p_{\mathrm{max}})+2C(n-1,p_{\mathrm{max}}-1)&\mathrm{~if~}n>p_{\mathrm{max}}
\end{array}
\end{equation}
\end{widetext}
Size classes are now described by
$C(n,p)=2^p\left(1+\tau\right)^{n-p}a_0$ for $0\leq p\leq n$ as long
as the generation index $p$ remains smaller than $p_{\mathrm{max}}$
and
$C(n,p_{\mathrm{max}})=2^{p_{\mathrm{max}}}\left[\left(1+\tau\right)^{n-p_{\mathrm{max}}+1}-1\right]a_0/\tau$
for $n\geq p_{\mathrm{max}}$. Whereas the filling of the size class
associated to ``atomic'' fragments was linear in $n$ without
inflation, it becomes here \emph{exponential}. Consequently, the
cumulative abundance, definitely slowed down, remains exponential in
$n$ for $n>p_{\mathrm{max}}$:
\begin{widetext}
\begin{equation}
    S_n=\left\{\left(1+\tau\right)^n\left[\frac{\left(\frac{2}{1+\tau}\right)^{p_{\mathrm{max}}}-1}{1-\tau}\right]+2^{p_{\mathrm{max}}}\frac{\left(1+\tau\right)^{n-p_{\mathrm{max}}+1}-1}{\tau}\right\}a_0
    \label{eq:cumAbundanceInflation}
\end{equation}
\end{widetext}
As long as the ``atomic limit'' is not reached, the cumulative
abundance exhibits a simpler form, namely:
\begin{equation}
    S_n=\left[2^{n+1}-\left(1+\tau\right)^{n+1}\right]\frac{a_0}{1-\tau}
    \label{eq:cumAbundanceInflationShortTime}
\end{equation}
that does not significantly differ from
Eq.~(\ref{eq:cumAbundanceStandard}). The time-invariant features of
the size distribution are nevertheless modified in two respects (see
Fig.~\ref{fig:inflationaryStandard}):
\begin{figure}[!ht]
\centering
\includegraphics[width=0.7\columnwidth]{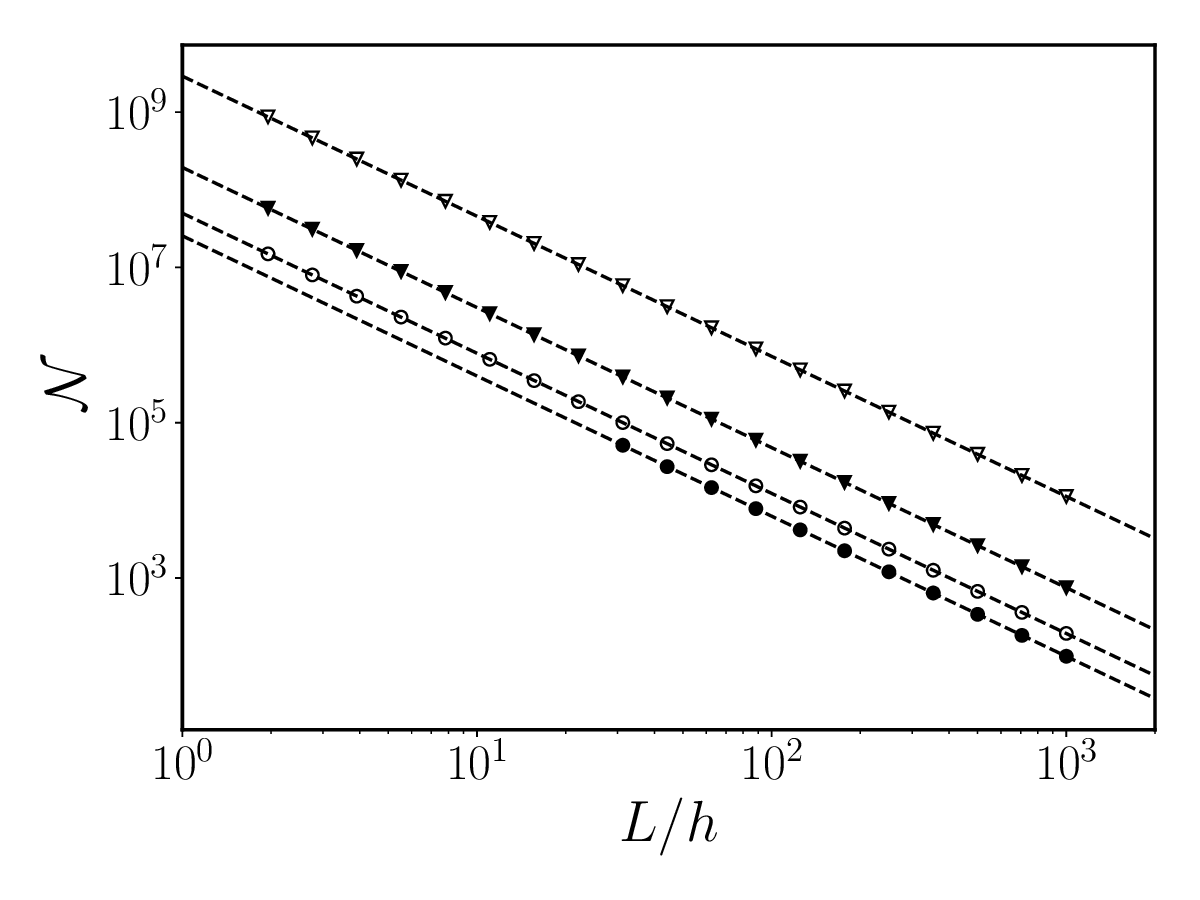}
\caption{Time-invariant features of the standard model, taking into
  account inflation at a 7\% yearly rate. Symbol keys and time
  indices: $\bullet$, $n=10$; $\circ$, $n=20$; $\blacktriangledown$,
  $n=40$; $\triangledown$, $n=80$. Superimposed dashed lines: fits to
  scaling law ${\cal N}\propto1/L^{\nu}$ with $\nu\approx1.8$}
\label{fig:inflationaryStandard}
\end{figure}
\begin{enumerate}
    \item Inflation spoils the strict time-invariant feature
      previously observed for the size distribution ${\cal N}(L)$;
    \item A (nearly) time-invariant behavior remains as far as
      \emph{scaling} is concerned, since ${\cal N}\propto1/L^{\nu}$,
      but $\nu$ does depend, albeit rather weakly, on the time index
      $n$, while being significantly smaller than $2$. Fitting data to
      a power law, an exponent $\nu$ close to $1.8$ is obtained for
      inflation $\tau=7\%$.
\end{enumerate}
\subsection{Sugar lump model}
\label{sugarLump}
\begin{figure}[!ht]
\centering
\includegraphics[width=0.7\columnwidth]{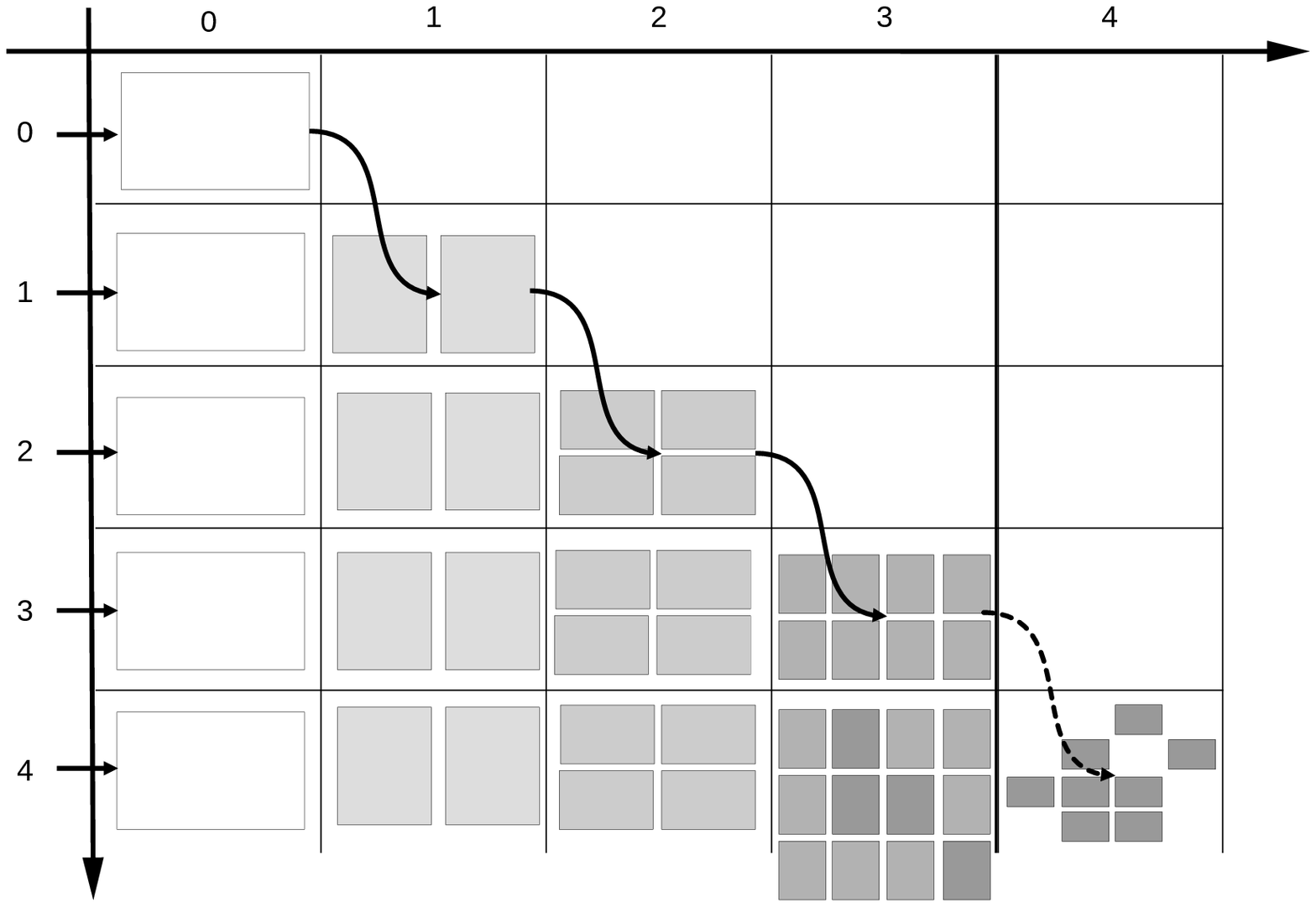}
\caption{Schematics of iterative size classes filling for the
  \emph{sugar lump} model. Horizontal axis, rightwards: size index
  $p$; Vertical axis, downwards: time index $n$. Short horizontal
  arrows stand for the feeding process. Full or dotted wavy arrows
  stand for a 100\%, or partially efficient fragmentation process. The
  shade of gray represents the \emph{weathering} (time spent in the
  ocean) of a given object. Note that the size threshold for
  fragmentation (vertical thick line between size classes 3 and 4)
  leads to a \emph{decoupling} between size and age}
\label{fig:remplissagecaseSeuil}
\end{figure}

Taking inspiration from the standard model,
Section~\ref{standardModel}, at each iteration the ocean is fed with
large parallelepipedic fragments of length $L$, width $\ell$ and
thickness $h$, where $h$ is much smaller than the other two dimensions
and length $L$ is, by convention, larger than width $\ell$. Some size
dispersity is introduced when populating the largest size class, by
randomly distributing $L$ in the interval $[0.9L_{\mathrm{init}},
  L_{\mathrm{init}}]$, and $\ell$ in $[0.7L_{\mathrm{init}},
  0.9L_{\mathrm{init}}]$, but $h$ is kept fixed. The \emph{number} of
objects feeding the system can be controlled at each iteration step,
and two simple limits have been investigated: Constant, or
exponentially-growing feeding rates, mimicking two variants of the
\emph{Standard model}, Sections~\ref{standardModel} and
\ref{inflationaryStandard}, respectively. Size-classes evenly sampling
(in \emph{logarithmic} scale) the full range of $L/h$, $[1,
  L_{\mathrm{init}}/h]$ are populated by sorting into the proper size
class the fragments present in the system. Except for the
$0^{\mathrm{th}}$, initialization step, these fragments are either
$0^{\mathrm{th}}$-generation fragments just introduced into the
system, obviously belonging to the largest size class, or
$g$-generation fragments ($g\geq1$) that have been ``weathered''
during the time step from step $n$ to step $n+1$ and then split, with
a $L$-dependent efficiency, into two smaller fragments. As explained
in Section~\ref{fragModelThreshold}, the splitting process, albeit
random, explicitly ensures the existence of an ``atomic'' limit:
Fragments belonging to the smallest size class cannot be fragmented
any further. As tentatively illustrated in
Fig.~\ref{fig:remplissagecaseSeuil}, a special feature of the model is
that \emph{generations} ($g$) and size-class ($p$) indices have to be
distinguished because, at contrast with the standard model, although
for a given fragment a ``weathering'' event ($n\rightarrow n+1$) is
always associated to an ``aging'' event ($g$ increased by one), it is
not always associated to populating one or two lower-size classes (and
simultaneously decreasing by 1 the abundance of the considered
size-class) because the splitting process is not $100$\% efficient.
Keeping track of abundances in terms of time ($n$), age ($g$) and size
($p$) being computationally demanding for exponentially-growing
populations, our simulations have been limited to, at most,
$n=g=40$. The number of distinct size classes has also been limited to
28, as this corresponds to the number of size-classes reported in
Ref.~\cite{cozar2014plastic}.
\bibliography{Biblio.bib}

\begin{thebibliography}{34}
\bibitem [{Notei()}]{Notei}%
  \BibitemOpen
  \bibinfo {note} {Provided that, of course, the depletion of very large
  objects that originates from the absence of feeding is
  disregarded.}\BibitemShut {Stop}%
\end{thebibliography}
\end{document}